\documentclass[preprint]{aastex}
\slugcomment{To appear in Astrophysical Journal Letters}
\shorttitle{Near-Infrared Spectra of ULIRGs}
\shortauthors{Murphy et al.}

\begin{document}

\title{Near-Infrared Spectra of Ultraluminous Infrared Galaxies}

\author{T. W. Murphy, Jr., B. T. Soifer\altaffilmark{1}, K. Matthews, J. R. Kiger\altaffilmark{2}}

\affil{Palomar Observatory, California Institute of Technology, 320-47, Pasadena,
CA 91125}

\email{tmurphy@mop.caltech.edu, bts@mop.caltech.edu, kym@caltech.edu}

\and

\author{L. Armus}

\affil{SIRTF Science Center, California Institute of Technology, 314-6, Pasadena,
CA 91125}

\altaffiltext{1}{Also at the SIRTF Science Center, California Institute of Technology, 314-6, Pasadena, CA 91125}

\altaffiltext{2}{Now at the Center for Space Research, Massachusetts Institute of Technology, Cambridge, MA 02139}

\begin{abstract}
Near infrared spectra with resolution \( \lambda /\Delta \lambda \approx  \)1100
in the rest wavelength range 1.8--2.2 \( \mu  \)m have been obtained for a
complete sample of 33 ultraluminous infrared galaxies. Of the 33 objects observed,
2 show evidence of a central AGN through either a broad Paschen~\( \alpha  \)
line or emission in the 1.963 \( \mu  \)m fine structure of {[}\ion{Si}{6}{]}.
In the median spectrum of the remaining 31 objects, the lines present are recombination
lines of Hydrogen, neutral Helium, vibration-rotation lines of H\( _{2} \),
and {[}\ion{Fe}{2}{]}. There is no indication of AGN activity in the median
spectrum, either through broad atomic recombination lines or high ionization
lines. No trends in luminosity are apparent when subsets of the 31 non-AGN ULIRGs
are binned by luminosity and median combined. When secondary nuclei exist in
ULIRGs, they typically have spectra very much like those seen in the primary
nuclei.
\end{abstract}


\section{Introduction}

From the initial realization that Ultraluminous Infrared Galaxies (hereafter
ULIRGs) represent a significant class of the highest luminosity objects in the
local universe \citep{bts87,san88}, a central question has been what powers
this luminosity. Various studies have attempted to address this question by
probing the centers of these highly dust enshrouded systems at wavelengths where
the opacity of dust is vastly reduced from that found at visible wavelengths,
ranging from X-ray \citep{reike,naka} to near-infrared \citep{carico,vx99} to
mid-infrared \citep{genzel,bts99}, as well as to radio wavelengths \citep{sll}.
These studies have generally found that direct detection of AGN (Active Galactic
Nucleus) properties in the nuclei of these systems is rare.

In this letter we summarize the results of a near-infrared spectroscopic survey
of 33 bright, nearby ULIRGs undertaken for the purpose of identifying sources
where the effects of dust might be hiding the presence of an AGN in the visible
spectra of ULIRGs. This is a natural explanation of the paucity of such spectra,
given that the ULIRGs are known to have extremely gas and dust rich nuclear
regions \citep{san88}. The reduced extinction in the near-infrared as compared
to the visible (i.e. \( A_{Pa\alpha }\sim 0.1A_{V} \), \citealp{extinct}) enables
extremely sensitive searches for characteristics of AGN hidden by large columns
of dust. This survey concentrates on searching for a high velocity component
to the strong Pa\( \alpha  \) line, as well as the presence of the very high
excitation {[}\ion{Si}{6}{]} line. The detailed results of this study will be
presented elsewhere \citep{twm00}.

\section{The Sample}

The complete sample of 33 ULIRGs for this spectroscopic study is a subset of
the sample of \citet{str90,str92}, which is a complete sample of galaxies selected
to have \( S_{\nu }> \)1.94 Jy at 60\( \mu  \)m and an infrared luminosity
of \( \ga 10^{12}L_{\odot } \). The selection criteria for the present sample
additionally require a declination \( \delta >-35^{\circ } \), and redshift
between \( 0.055<z<0.108 \). This last constraint allows observations of both
Pa\( \alpha  \) and Br\( \gamma  \) in the 2\( \mu  \)m atmospheric window.
The list of sample galaxies can be read directly out of \citet{twm96} with the
above redshift constraint imposed. Two of the 35 galaxies from this sublist
have been removed from the present sample: IRAS~21396\( + \)3623 is actually
at a redshift of \( z=0.1493 \), and IRAS~19297\( - \)0406, at galactic latitude
\( b\sim -11^{\circ } \), does not appear to be properly identified as a ULIRG.

\section{Observations and Data Reduction}

Observations were obtained with the Palomar Longslit Infrared Spectrograph \citep{lark} on
observing runs over the period 1996 January to 1997 December. In all cases the
object was observed in at least 2 grating settings, and for about 90\% of the
objects 3 grating settings were employed. Each grating setting covers \( \sim 0.12\mu  \)m
with a scale of 0.0006 \( \mu  \)m~pixel\( ^{-1} \). The slit width of 4 pixels,
or 0\farcs 67 corresponds to a spectral resolution of \( R\equiv \lambda /\Delta \lambda \approx 1100 \),
corresponding to \( \sim  \)280 km~s\( ^{-1} \). The grating settings were
chosen to cover the Pa\( \alpha  \) line at 1.8751\( \mu  \)m, the suite of
Br\( \delta  \), H\( _{2} \) 1--0~S(3) and {[}\ion{Si}{6}{]} lines centered
at 1.954\( \mu  \)m, and if a third setting was used, the H\( _{2} \) 1--0~S(1)
and Br\( \gamma  \) lines centered at 2.14\( \mu  \)m. The slit was generally
rotated to coincide with the major axis of the ULIRG, where evident, or such
that spectra of a secondary nucleus was simultaneously obtained when possible.

Typically, integration times were 1800 s at each grating setting, with the object
alternately displaced along the slit such that sky integrations were obtained
simultaneously. Wavelength calibration was provided by a combination of OH airglow
lines and arc lamp spectra. Atmospheric transmission is compensated by observing
the nearly featureless spectra of G \textsc{V} stars, and the spectrum divided
by a Planck curve of appropriate temperature. The Br\( \gamma  \) absorption
feature in the G star is removed by interpolation. Other minor absorption features
present in the G star calibration spectrum are removed using a template spectrum
from \citet{templates}. Spectral extractions, centered on the primary nucleus
of each ULIRG, were chosen to match the seeing at the time of observation, typically
0\farcs 8--1\farcs 2. Extractions on the secondary nuclei, when applicable,
were performed in the same manner. A more detailed description of the observations
and data reduction procedures are deferred to \citet{twm00}, in which the individual
spectra will also be presented.

\section{Results}

The most striking property of the collection of ULIRG spectra is that, with
a few notable exceptions, there exists remarkable similarity among the vast
majority (31 of 33) of spectra of galaxies in the sample. This similarity prompted
us to combine all the similar spectra into a single, representative composite
ULIRG spectrum. It is our belief that this composite spectrum provides an accurate
representation of the typical ULIRG spectrum.

Figure~\ref{fig1} shows the median of spectra from 31 of the ULIRGs normalized
to their continuum level, and shifted in wavelength corresponding to the redshift
of Pa\( \alpha  \). The objects selected to be included in the combination
were chosen to show no obvious evidence for broad lines or high excitation lines
based on inspection of the individual spectra. Combining the spectra facilitates
identification of any faint features that might be commonly present but with
low signal to noise ratio in individual spectra. An average spectrum was also
generated, the features of which are indistinguishable from those in the median
spectrum. Specific properties and identifications of the lines in the median
spectrum are discussed in Section~\ref{lineprops}.

Figure~\ref{fig2} demonstrates the range of line strengths seen in the ULIRG
sample in the form of equivalent widths of the Pa\( \alpha  \) and H\( _{2} \)
1--0~S(3) lines. Most of the ULIRGs lie within roughly one bin-width of the
median value, as displayed in Figure~\ref{fig2}, though there are a number
of outliers with significantly higher equivalent widths than the median value.
We do not find any correlation between the Pa\( \alpha  \) and H\( _{2} \)
equivalent widths.

A median spectrum of 13 secondary ULIRG nuclei has also been constructed for
the purpose of comparing the properties of primary and secondary nuclei. While
two of the secondary nuclei exhibit featureless spectra, the median spectrum
shows that secondary nuclei tend to be spectroscopically similar to the primary
ULIRG nuclei. In general, all of the spectral features visible in the primary
nuclei (i.e. Figure~\ref{fig1}), with the exception of the \ion{He}{1} line,
are visible in the median spectrum composed of secondary ULIRG nuclei, though
with lower equivalent widths.

\subsection{Median Spectrum Line Properties\label{lineprops}}

\subsubsection{Atomic Recombination Lines}

As can be seen in Figure~\ref{fig1}, Pa\( \alpha  \) emission dominates the
near-infrared spectrum of the typical ULIRG. Other recombination lines of hydrogen,
specifically Br\( \gamma  \), Br\( \delta  \), and Br\( \epsilon  \) are
also readily seen in the median spectrum. The extinction to the line emitting
region can be crudely estimated from the ratio of the Pa\( \alpha  \) and Br\( \gamma  \)
lines. The reddening derived in this way from the median spectrum is \( A_{V}=5 \)--10
mag. There is some evidence that the extinction is higher in the secondary ULIRG
nuclei.

The weak line at 1.869\( \mu  \)m are identified with a pair of \ion{He}{1}
recombination lines at 1.8686\( \mu  \)m and 1.8697\( \mu  \)m, with a centroid
at 1.8689\( \mu  \)m. We have found that this line is comparable in strength
to the \ion{He}{1} line at 2.058\( \mu  \)m. Because of the small separation
between the Pa\( \alpha  \) line and the \ion{He}{1} line, at lower spectral
resolution this line might be confused with a blue wing of the Pa\( \alpha  \)
line.

\subsubsection{Molecular Hydrogen Lines}

Vibration-rotation lines of H\( _{2} \) are also prominent in the median spectrum.
The H\( _{2} \) 1--0 lines S(1), S(3), S(4), and S(5) are clearly present in
the median spectrum and the S(2) line is evident at low significance. There
is marginal evidence for the H\( _{2} \) 2--1~S(2) and S(4) lines in the median
spectrum, but these are at very low significance. The observed H\( _{2} \)
spectrum is consistent with purely thermal excitation at a temperature of \( T=T_{vib}=T_{rot}\sim  \)2500
K. At this temperature, contributions to the spectrum from H\( _{2} \) 6--4~O(5)
and 7--5~O(3) at \( \lambda \lambda  \)1.8665, 1.8721, potential contributors
to the blue wing of Pa\( \alpha  \) \citep[e.g.,][]{vx99}, are of negligible
significance. The appearance of line emission at the position of the 2--1 S(4)
line is inconsistent with the apparent weakness of the 2--1~S(2) line, suggesting
the presence of {[}\ion{Fe}{2}{]} at 2.0024\( \mu  \)m. This possibility is
further discussed below.

At a spectral resolution of 280 km~s\( ^{-1} \), the \ion{H}{1} lines appear
to be spectrally unresolved in the median spectrum of Figure~\ref{fig1}. The
H\( _{2} \) lines, on the other hand, appear to have a broader spectral profile,
especially evident near the bases of these lines. Random velocity offsets between
the \ion{H}{1} and H\( _{2} \) line emission would indeed tend to broaden the
H\( _{2} \) lines when using Pa\( \alpha  \) as the redshift reference for
the combination. However, when the H\( _{2} \) lines themselves are used as
the redshift reference, the H\( _{2} \) profiles are seemingly unaffected,
with no apparent broadening of the Pa\( \alpha  \) profile.

\subsubsection{Iron Lines}

A faint line appears in the median spectrum at 1.967\( \mu  \)m, close to the
wavelength of {[}\ion{Si}{6}{]} at 1.963\( \mu  \)m. We do not identify this
line with the {[}\ion{Si}{6}{]} line for two reasons. First, in two objects
we directly detect the {[}\ion{Si}{6}{]} line correctly centered at 1.963\( \mu  \)m
(see below). In addition, there is a matching line from {[}\ion{Fe}{2}{]} at
1.967\( \mu  \)m. Such a feature is consistent with the known strength of {[}\ion{Fe}{2}{]}
lines in starburst galaxies of lower luminosity. This identification would imply
that the {[}\ion{Fe}{2}{]} 1.644\( \mu  \)m and 1.258\( \mu  \)m lines should
be present at much greater strength in these spectra. While such observations
are limited there is evidence, e.g., \citet{vx97} that such strong Fe lines
are common in ULIRGs. The presence of the 1.967\( \mu  \)m {[}\ion{Fe}{2}{]}
line also supports the existence of the suspected {[}\ion{Fe}{2}{]} line at
2.0024\( \mu  \)m, coincident with the H\( _{2} \) 2--1~S(4) line. The 2.0024\( \mu  \)m
{[}\ion{Fe}{2}{]} line is expected to be two to three times weaker than the
line at 1.967\( \mu  \)m, which is consistent with the observed low-level line
at 2.003\( \mu  \)m being a blend of {[}\ion{Fe}{2}{]} and H\( _{2} \) lines.

\subsection{ULIRGs with AGN Properties}

Figure~\ref{fig3} shows spectra of the two objects in the sample where evidence
of AGN associated spectral features is clearly present in the individual object
spectra. While the spectra of these two galaxies have characteristics not found
in the median spectrum, these spectra are still quite similar in appearance
to the median spectrum of Figure~\ref{fig1}.

The near-infrared spectrum of IRAS~08311\( - \)2459 (\( z=0.1006 \)), presented
in Figure~\ref{fig3}, shows a strong, and slightly broad {[}\ion{Si}{6}{]}
line along with the \ion{H}{1} recombination lines and H\( _{2} \) lines. The
strength of the {[}\ion{Si}{6}{]} line is quite unusual for ULIRGs in general,
being roughly equal in strength to the H\( _{2} \) 1--0~S(3) line. The presence
of the {[}\ion{Si}{6}{]} line directly demonstrates the existence of high energy
photons since the ionization potential of Si\( ^{4+} \) is 167 eV. IRAS~08311\( - \)2459
also shows the suggestion of a broad Pa\( \alpha  \) base with an asymmetric
blue wing that is not consistent with being a \ion{He}{1} recombination line.
This appears to be a high velocity component of the \ion{H}{1} line.

The near-infrared spectrum of IRAS~15462\( - \)0450 (\( z=0.1003 \)), presented
in Figure~\ref{fig3}, shows a broad component of Pa\( \alpha  \) in addition
to the strong narrow component of this line, as well as evidence for the \ion{He}{1} line.
There is no evidence for the {[}\ion{Si}{6}{]} line in the spectrum, though
the H\( _{2} \) 1--0~S(3) line is clearly present.

For comparison, a median quasar spectrum is included in Figure~\ref{fig3}.
This spectrum is composed from ten quasars ranging in redshift from \( z=0.089 \)--0.182,
observed with the same instrument and in the same manner as that described above
for the ULIRGs. The median quasar spectrum is dominated by broad Pa\( \alpha  \)
emission, though a velocity-broadened blend of Br\( \delta  \), H\( _{2} \)
1--0~S(3), and {[}\ion{Si}{6}{]} can also be discerned. The {[}\ion{Si}{6}{]}
line appears to be comparable in strength to the H\( _{2} \) 1--0~S(3) line,
as is also the case for IRAS~08311\( - \)2459. A more detailed account of the
quasar spectra will accompany the paper describing the individual ULIRG spectra
\citep{twm00}.

\section{Discussion}

The most direct result from this work is that the great majority of ULIRGs (31
of 33) show no indication of AGN activity when viewed at 2\( \mu  \)m with
relatively high spectral resolution and sensitivity. Of the 33 objects observed,
90\% have luminosities \( L<2\times 10^{12}L_{\odot } \) and of these nearly
all (29/30) lack spectroscopic evidence for broad line regions or high energy
photons. In the 3 objects with \( L_{ir}\geq 2\times 10^{12}L_{\odot } \),
one shows spectroscopic evidence of AGN activity. While lacking the numbers
to make a statistically significant argument, this fraction is consistent with
findings in previous near-infrared spectroscopic surveys, even though our resolution
and sensitivity is higher than the previous studies \citep{vx97,vx99}.

Where evidence for AGN activity is present, it is readily detectable in the
individual spectrum, and spectroscopic evidence for the AGN activity is also
present in the visible light spectrum of the object. IRAS~15462\( - \)0450
is classified as a Seyfert~1 system based on its visible spectrum, while our
unpublished observations of IRAS~08311\( - \)2459 show that it has a Seyfert~2
visible spectrum. \citet{lutz} also report that ULIRGs showing infrared signatures
of the AGN phenomenon are usually classified as active galaxies based on their
visible light spectra. The converse statement---that all ULIRGs showing no evidence
for AGN activity in their near-infrared spectra are optically classified as
non-active galaxies---seems also to be supported. Of the 16 non-AGN ULIRGs in
the present sample with available optical spectroscopic identifications, only
one (IRAS~14394\( + \)5332) is characterized as a Seyfert galaxy based on a
marginally high {[}\ion{O}{3}{]}/H\( \beta  \) ratio \citep{kvs}. While this
galaxy does exhibit a blue wing on the \ion{H}{1} and H\( _{2} \) lines, there
is no corresponding emission redward of the unresolved narrow line component.
We therefore do not associate this velocity structure with a massive compact
central source, but rather with organized gas motion, such as might be found
in outflow phenomena.

The median spectrum of the ULIRGs that do not show evidence for AGN activity
presents an important new result, i.e. there are not faint broad or high excitation
lines present in the spectra that are not detected in the individual spectra.
The relative \ion{H}{1} and H\( _{2} \) line strengths in the AGN and median
spectra are quite comparable, while the limits on the strength of a broad component
of Pa\( \alpha  \) or {[}\ion{Si}{6}{]} 1.963\( \mu  \)m are at least an order
of magnitude less in the median spectrum than in the individual sources with
AGNs clearly evident. Furthermore, no luminosity dependent spectroscopic trends
are seen in the 31 non-AGN ULIRGs when the spectra are divided into subsets
and median-combined in separate luminosity bins. In particular, there is no
more evidence for low-level broad Pa\( \alpha  \) in the higher luminosity
ULIRGs than there is in the lower luminosity sets or in the collection as a
whole. These findings can be explained in several ways. Most obviously it could
be that there are simply not AGN associated with the ULIRGs where such features
are not obvious. This would imply that for \( L_{ir}<2\times 10^{12}L_{\odot } \),
only a very small fraction of ULIRGs are powered by AGNs. Presumably the power
source in these cases is star formation in extremely compact star forming regions.
Alternatively, if all these sources harbor an AGN it could be that the fraction
of sources with visible spectroscopic evidence for AGN activity is a measure
of geometric effects, i.e. there are highly attenuating dust tori with \( A_{Pa\alpha }> \)2.5
mag covering \textgreater{} 90\% of the sky from the central source. Other more
carefully contrived configurations of dust are also possible.

While we cannot conclusively argue one way or another, the highly sensitive
survey we have conducted for dust enshrouded AGN, combined with other such surveys
at other wavelength is providing mounting evidence that star formation is powering
the vast majority of ULIRGs with \( L_{ir}<2\times 10^{12}L_{\odot } \).

\acknowledgements

We thank Michael Strauss for his role in defining the sample of ULIRGs, and
for participating in the early stages of the Caltech effort in studying ULIRGs.
We also thank Gerry Neugebauer for helpful discussions. Many persons accompanied
us on the observing runs to Palomar, most notably Rob Knop and James Larkin,
both of whom provided expert tutelage on the use of the spectrograph, and on
methods of data reduction. We thank the night assistants at Palomar, Jaun Carasco,
Rick Burruss, Skip Staples, and Karl Dunscombe. for their assistance in the
observations. T.W.M. is supported by the NASA Graduate Student Researchers Program,
and the Lewis Kingsley Foundation. This research is supported by a grant from
the National Science Foundation.

\clearpage

\figcaption[FileName]{\label{fig1}Median ULIRG spectrum, composed of 31 ULIRGs showing no evidence
for AGN activity in their individual spectra. Pa\protect\( \alpha \protect \)
dominates the \protect\( K\protect \) band spectrum, with strong contributions
from molecular hydrogen and from other atomic recombination lines. There is
no evidence for broad Pa\protect\( \alpha \protect \) emission or for the {[}\ion{Si}{6}{]}
line in the combined spectrum. A 3\protect\( \sigma \protect \) envelope about
the continuum level is provided for reference. The median spectrum is not displayed
for wavelength regions in which fewer than five ULIRG spectra are available
for combination.}

\figcaption[FileName]{\label{fig2}Distribution of equivalent widths for the Pa\protect\( \alpha \protect \)
and H\protect\( _{2}\protect \) 1--0~S(3) lines, representing the ULIRGs that
comprise the median spectrum in Figure~\ref{fig1}. The median and average equivalent
width values are marked in each of the histograms.}

\figcaption[FileName]{\label{fig3}Individual spectra of the two ULIRGs in the sample showing some
signs of AGN activity. IRAS~08311\protect\( -\protect \)2459 has a strong {[}\ion{Si}{6}{]}
detection plus a low-level broad blue wing on the Pa\protect\( \alpha \protect \)
line. IRAS~15462\protect\( -\protect \)0450 shows a broad base on the Pa\protect\( \alpha \protect \)
line, though no evidence for {[}\ion{Si}{6}{]} emission. The median quasar spectrum
is provided as a reference, showing both the broad Pa\protect\( \alpha \protect \)
and {[}\ion{Si}{6}{]} emission features. The line at \protect\( \lambda 1.914\mu \protect \)m
in the spectrum of IRAS~15462\protect\( -\protect \)0450 is a residual from
Mg absorption in the G star atmospheric calibrator. The zero flux density ``floors''
for the three spectra lie at the values 0.4, 0.15, and 0.0 on the vertical axis,
ordering the spectra from top to bottom.}

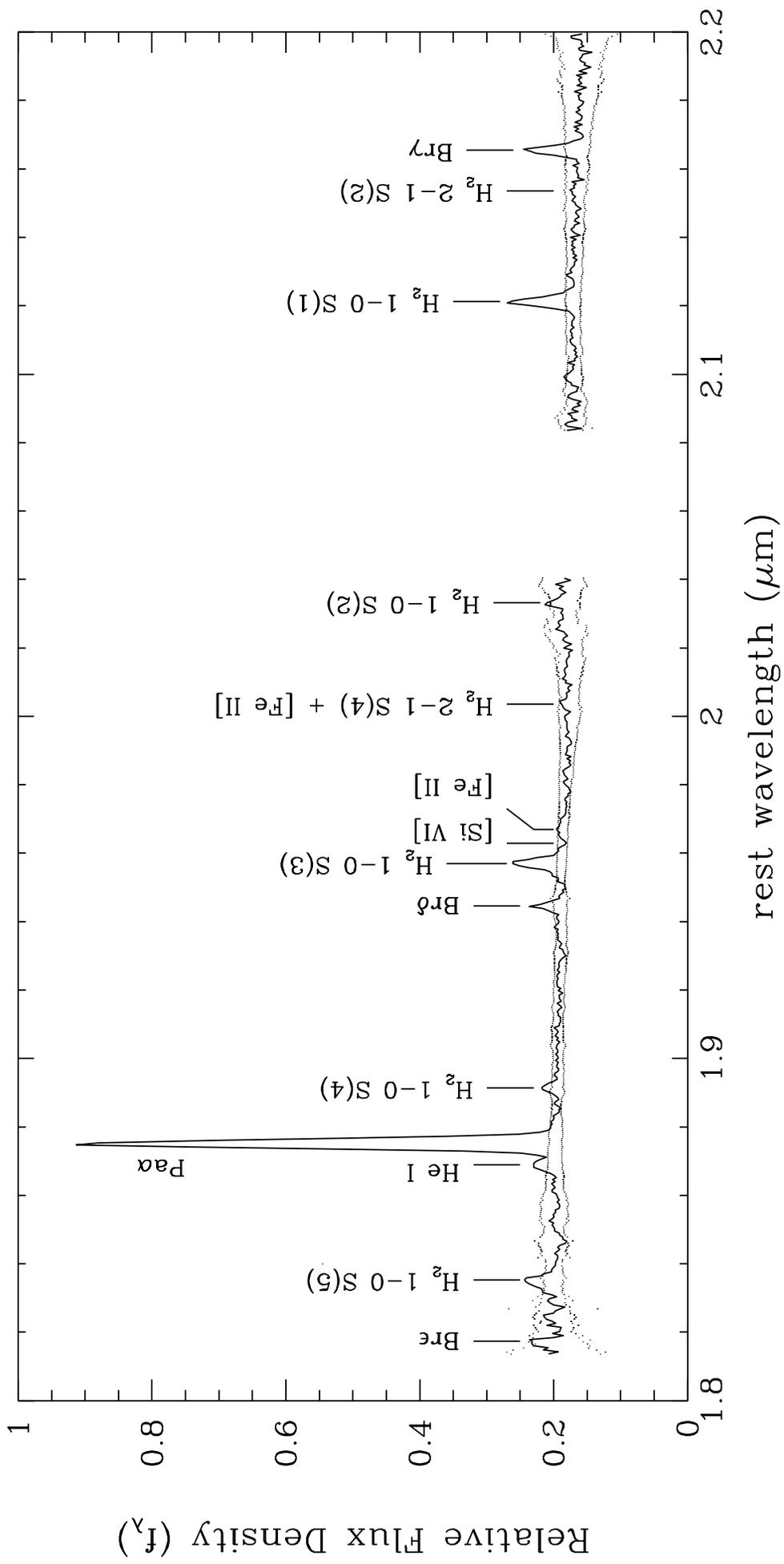
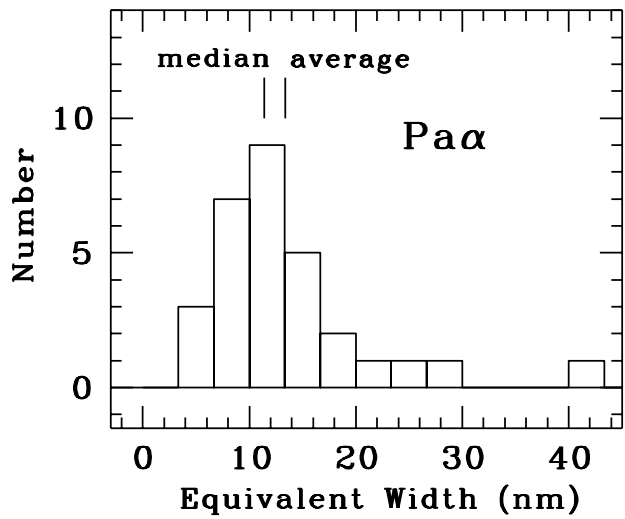
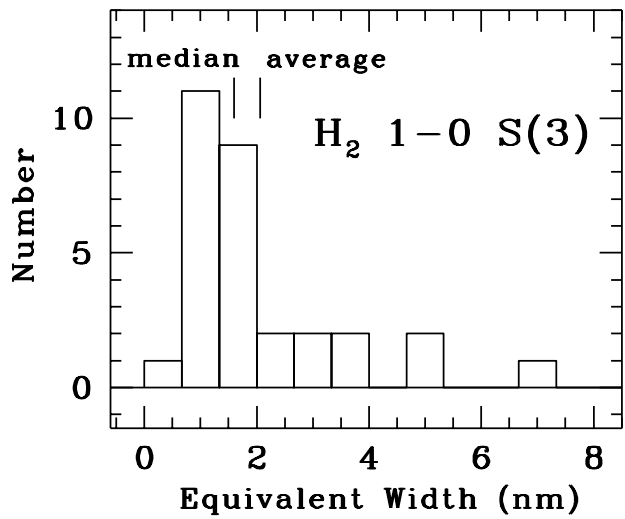
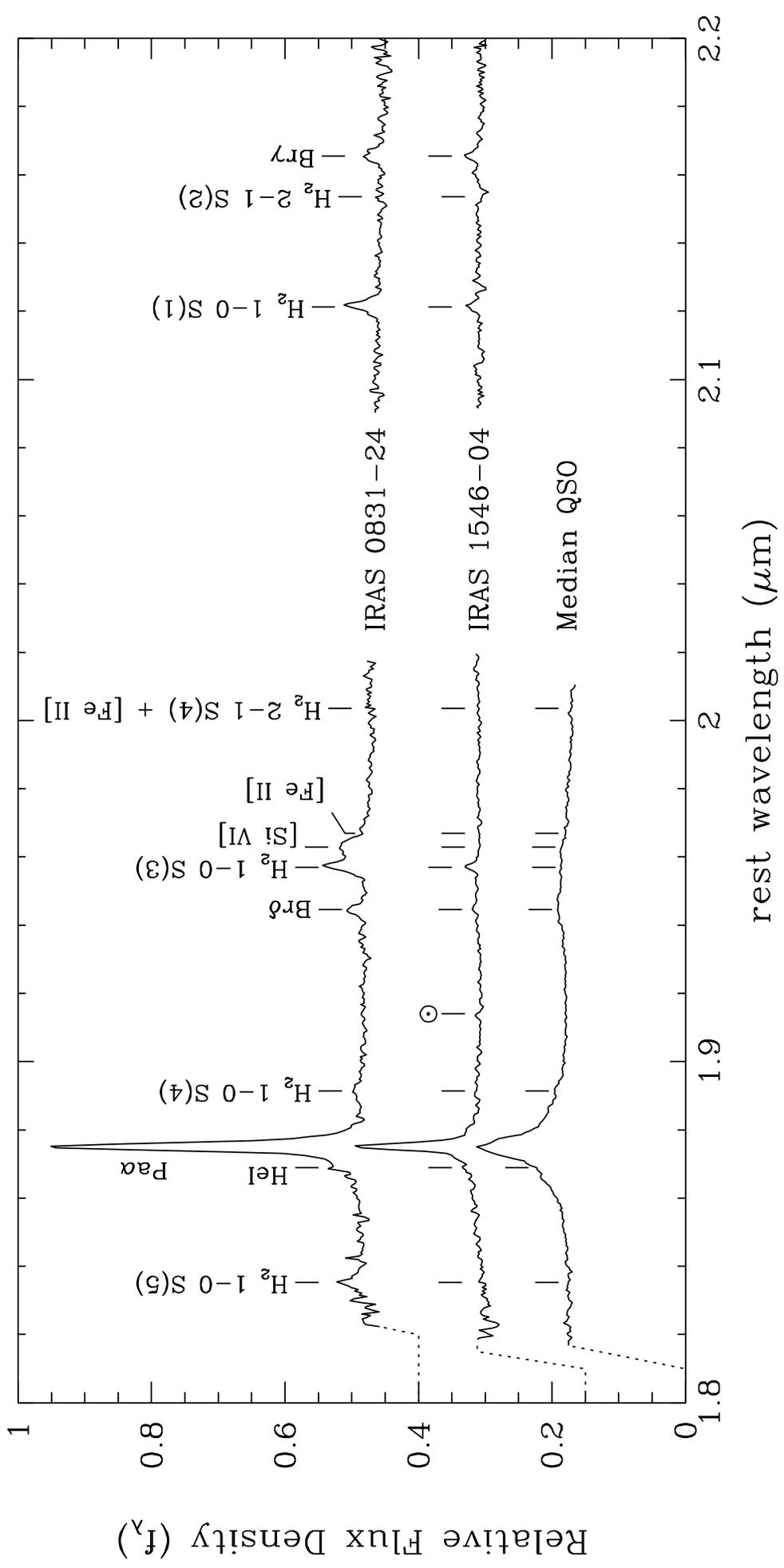
\end{document}